\documentclass[pdf,reqno]{amsart}
\usepackage{amssymb,amsmath,amscd}
\theoremstyle{plain}
\numberwithin{equation}{section}
\newtheorem{theorem}{Theorem}[section]

\theoremstyle{definition}
\newtheorem{definition}[theorem]{Definition}

\newtheorem{remark}[theorem]{Remark}
\newtheorem{example}[theorem]{Example}

\usepackage{hyperref}
\usepackage{amssymb, amsmath}
\newcommand{\bE}{{\mathbb E}}
\newcommand{\E}{{\bE}} 
\newcommand{\cH}{\mathcal{H}}
\newcommand{\cE}{\mathcal{E}}

\newcommand{\la}{\langle}
\newcommand{\ra}{\rangle}

\newcommand{\OO}{\mathop{\rm O{}}\nolimits}

\newcommand{\U}{\mathop{\rm U{}}\nolimits}

\newcommand{\Mot}{{\rm Mot}}

\newcommand{\g}{{\mathfrak g}}
\newcommand{\fh}{{\mathfrak h}}
\newcommand{\fq}{{\mathfrak q}}
\newcommand{\vphi}{\varphi}
\newcommand{\C}{{\mathbb C}}
\newcommand{\K}{{\mathbb K}}
\newcommand{\R}{{\mathbb R}}
\newcommand{\N}{{\mathbb N}}
\newcommand{\1}{\mathbf{1}}
\newcommand{\id}{\mathop{{\rm id}}\nolimits}
\newcommand{\sgn}{\mathop{{\rm sgn}}\nolimits}
\newcommand{\res}{\vert}
\newcommand{\subeq}{\subseteq}

\begin{document}
\title{Reflection positive affine actions and stochastic processes} 

\author{P. E. T. Jorgensen} 

\address{Department of Mathematics, The University of Iowa \\ 
Iowa City, IA 52242, USA \\ 
E-mail: palle-jorgensen@uiowa.edu} 

\author{K.-H. Neeb$^*$} 

\address{Department Mathematik, FAU Erlangen-N\"urnberg \\ 
Cauerstrasse 11, 91058-Erlangen, Germany\\ 
$^*$E-mail: neeb@math.fau.de}

\author{G. \'Olafsson} 

\address{Department of mathematics, Louisiana State University \\  
Baton Rouge, LA 70803, USA \\ 
E-mail: olafsson@math.lsu.edu}

\begin{abstract}
In this note we continue our investigations of the representation 
theoretic aspects of reflection positivity, 
also called Osterwalder--Schrader positivity. 
We explain how this concept relates to affine isometric actions 
on real Hilbert spaces and how this is connected with Gaussian processes
with stationary increments.
\end{abstract}


\maketitle

\section{Introduction} 
In this note we continue our investigations of the mathematical 
foundations of \textit{reflection positivity}, also called Osterwalder--Schrader positivity, a basic concept in constructive quantum 
field theory \cite{GJ81, Kl77, KL82, JO00, JR07}. 
It  
arises as a requirement on the euclidean side to establish a 
duality between euclidean and relativistic quantum field theories \cite{OS73}. 
It is closely related to ``Wick rotations'' or 
``analytic  continuation'' in the time variable 
from the real  to the imaginary axis. 

The underlying concept is that of a 
{\it reflection positive Hilbert space}, introduced in \cite{NO14}. 
This is a triple $(\cE,\cE_+,\theta)$, 
where $\cE$ is a Hilbert space, $\theta : \cE \to \cE$ is a unitary involution
and $\cE_+$ is a closed subspace of $\cE$ which is $\theta$-positive in the sense that 
the hermitian form $\langle \theta u,v\rangle$ is
positive semidefinite on $\cE_+$.\begin{footnote}  {As customary in physics, 
we follow the convenient that the inner product of a complex
 Hilbert space is linear in the second argument.}
\end{footnote}
We write $\widehat\cE$ for the corresponding Hilbert space 
and $q \colon \cE_+ \to \widehat\cE$ for the canonical map. 

To see how this relates to group representations, let us call a triple 
$(G,S,\tau)$ a {\it symmetric Lie semigroup} if $G$ is a Lie group, 
$\tau$ is an involutive automorphism of $G$ and $S \subeq G$ a 
unital subsemigroup (with dense interior) 
invariant under the involution $s \mapsto s^\sharp := \tau(s)^{-1}$.  
The Lie algebra 
$\g$ of $G$ decomposes into $\tau$-eigenspaces $\g = \fh \oplus \fq$ 
and we obtain the {\it Cartan dual Lie algebra} $\g^c=\fh\oplus i\fq$. $G^c$ then
stands for a Lie group with Lie algebra $\g^c$, often taken simply connected to make
it unique. The prototypical pair $(G,G^c)$ consists of the euclidean 
motion group $E(d) = \R^d \rtimes \OO_d(\R)$ and the simply connected covering of the 
orthochronous 
Poincar\'e group $P(d)^\uparrow =  \R^d \rtimes \OO_{1,d-1}(\R)^\uparrow$. 

If $(G,H,\tau)$ is a symmetric Lie group and
$(\cE,\cE_+,\theta)$ a reflection positive Hilbert space, then we say that 
a unitary representation $\pi \colon G \to \U(\cE)$ is {\it reflection positive 
with respect to $(G,S,\tau)$} if 
\[ \pi(\tau(g)) = \theta \pi(g)\theta \quad \mbox { for } g \in G 
\quad \mbox{ and } \quad \pi (S)\cE_+\subeq\cE_+.\] 

If $(\pi,\cE)$ is a reflection positive representation of $G$ on 
$(\cE,\cE_+, \theta)$, then $\widehat\pi(s)q(v) := q(\pi(s)v)$ 
defines a representation
$(\widehat\pi, \widehat\cE)$ of the involutive semigroup $(S,\sharp)$ by contractions 
(Lemma 1.4 in Ref.~\cite{NO14} or Ref.~\cite{JO00}). 
However, we would like to have a 
unitary representation $\pi^c$ of the simply connected Lie group $G^c$ with 
Lie algebra $\g^c$ on $\widehat\cE$ whose derived representation 
is compatible with the representation of~$S$. 
If such a representation exists, then we call 
$(\pi, \cE)$ a {\it euclidean realization} 
of the representation $(\pi^c,\widehat\cE)$ of $G^c$. 
Sufficient conditions for the existence of $\pi^c$ have been developed in 
\cite{MNO14} (see also \cite{KL82,LM75}). 

The main new aspects introduced in this short note is a notion of 
reflection positivity for affine actions of $G$ on a real Hilbert space. 
Here $\cE_+$ is generated by the $S$-orbit of the origin. On the level 
of positive definite functions, this leads to the notion of a reflection 
negative function. For $(G,S,\tau) = (\R,\R_+,-\id_\R)$, 
reflection negative functions $\psi$ are easily determined because 
reflection negativity is equivalent to $\psi\res_{(0,\infty)}$ being a Bernstein 
function. We conclude this note with a brief discussion of 
connections with some stochastic processes on the real line. 
Proofs, more details, and more complete results will appear in \cite{JNO16}.

\section{From kernels to reflection positive representations} 

Since our discussion is based on positive definite kernels, 
the corresponding Gaussian processes and the associated Hilbert spaces, 
we first recall the pertinent definitions. 

\begin{definition}   \label{def:8.1.1} 
(a) Let $X$ be a set. A kernel $Q \colon X \times X \to \C$ is called 
{\it hermitian} if $Q(x,y) = \overline{Q(y,x)}$. 
A hermitian kernel $Q$ is called
 {\it
positive}, respectively {\it negative}, {\it definite}, if for 
$x_1, \ldots, x_n \in X, c_1, \ldots, c_n \in \C$, we have 
$\sum_{j,k=1}^n c_j \overline{c_k} Q(x_j, x_k) \geq 0$, respectively if in addition
$\sum c_j=0$, we have  $\sum_{j,k=1}^n c_j \overline{c_k} Q(x_j, x_k) \leq 0$
\cite{BCR84}. 

(b) If $G$ is a group, then a function $\vphi \colon G \to \C$ is called {\it positive (negative) definite} if the kernel $(\vphi(gh^{-1}))_{g,h \in G}$ is positive (negative) definite. 
More generally, if $(S,*)$ is an involutive semigroup, then $\vphi \colon S \to \C$ is 
called {\it positive (negative) definite} if the kernel $(\vphi(st^*))_{s,t\in S}$ 
is positive (negative) definite. 
\end{definition} 

\begin{remark} \label{rem:kerspace} 
Let $X$ be a set, $K \colon X \times X \to \C$ be a positive definite 
kernel and $\cE = \cH_K \subeq \C^X$ the corresponding {\it reproducing kernel 
Hilbert space}. This is the unique Hilbert subspace of $\C^X$ on which all 
point evaluations $f \mapsto f(x)$ are continuous and given by 
$f(x) = \la K_x, f \ra$ for $K_x(y) = K(x,y)$. 
Then the map $\gamma \colon X \to \cH_K, \gamma(x) = K_x$ has total range 
and satisfies $K(x,y) = \la \gamma(x),\gamma(y)\ra$. The latter property 
determines the pair $(\gamma, \cH_K)$ up to unitary equivalence. 
\end{remark}

\begin{example} \label{ex:1.3} 
(a) Suppose that $K \colon X \times X \to \C$ is a positive definite kernel 
and $\tau \colon X \to X$ is an involution 
leaving $K$ invariant and that $X_+ \subeq X$ is a subset with the property that the 
kernel $K^\tau(x,y) := K(\tau x, y)$ is also positive definite on $X_+$. 
Then the closed subspace $\cE_+ \subeq \cE := \cH_K$ generated 
by $(K_x)_{x \in X_+}$ is $\theta$-positive for 
$(\theta f)(x) := f(\tau x)$. We thus obtain a 
reflection positive Hilbert space $(\cE,\cE_+,\theta)$. 
We call such kernels $K$ {\it reflection positive} with respect to $(X,X_+, \tau)$. 

(b) If $(\cE, \cE_+,\theta)$ is a reflection positive Hilbert space, 
then the scalar product defines a reflection positive kernel 
$K(v,w) := \la v, w \ra$ with respect to $(\cE,\cE_+,\theta)$. In this sense all 
reflection positive Hilbert spaces can be obtained in the context of (a), which 
provides a ``non-linear'' setting for reflection positive Hilbert spaces. 
\end{example} 

For symmetric Lie semigroups $(G,S,\tau)$, 
we obtain natural examples of reflection positive kernels: 
A function $\vphi \colon G \to \C$ is 
called {\it reflection positive} if the kernel 
$K(x,y) := \vphi(xy^{-1})$ is reflection positive with respect to $(G,S,\tau)$ 
in the sense of Example~\ref{ex:1.3}. 
These are two simultaneous positivity conditions, namely that the kernel 
$\vphi(gh^{-1}))_{g,h \in G}$ is positive definite on $G$ and that the kernel 
$\vphi(st^\sharp))_{s,t \in S}$ is positive definite on $S$.  Here $X=G$ and
$X_+=S$.

Prototypical examples are the functions 
$\vphi(t) = e^{-\lambda|t|}$, $\lambda \geq 0$, for $(\R,\R_+, -\id_\R)$
with $\widehat\cE$ one dimensional. For this triple   every continuous reflection positive 
function has an integral representation 
$\vphi(t)= \int_0^\infty e^{-\lambda|t|}\, d\mu(\lambda)$ for a positive measure 
$\mu$ on $[0,\infty)$ (see Cor.~3.3 in Ref.~\cite{NO14}). 
In Ref.~\cite{NO14} we also discuss generalizations of this concept to distributions 
and obtain integral representations for the case where $G$ is abelian. 

\section{Second quantization and Gaussian processes} 

\begin{definition} \cite{Hid80}  Let $T$ be a set 
and $\K = \R$ or $\C$. 
A $\K$-valued stochastic process $(X_t)_{t \in T}$ 
is said to be {\it Gaussian} if, for all finite subsets 
$F \subeq T$, the corresponding distribution of the random vector 
$X_F= (X_t)_{t \in F}$ with values in $\K^F$ is Gaussian. 
\end{definition}

\begin{definition} (Second quantization) \label{def:3.5}
(a) For a real Hilbert space $\cH$, we write 
$\cH^*$ for its algebraic dual, i.e., the set of all 
(not necessarily continuous) linear functionals 
$\cH\to \R$, continuous or not. Let $\Gamma(\cH) := L^2(\cH^*, \gamma,\C)$, 
where $(\cH^*,\Sigma, \gamma)$ is the canonical Gaussian measure space on $\cH^*$ 
(Ex.~4.3 in Ref.~\cite{JNO15}; see also Ref.~\cite{Si74}). 
Here $\Sigma$ is the smallest $\sigma$-algebra 
for which all evaluations $\phi(v) \colon \cH^* \to \R, \alpha\mapsto \alpha(v)$, $v \in \cH$, 
are measurable and the probability measure $\gamma$ is determined uniquely by 
\begin{equation}
  \label{eq:a}
\E(e^{i\phi(v)}) = e^{-\|v\|^2/2} \quad \mbox{ for } \quad v \in \cH.
\end{equation}
Considering the $\phi(v)$ as random variables, 
we thus obtain the {\it canonical Gaussian process} $(\phi(v))_{v \in \cH}$ over~$\cH$. 
It satisfies 
\[ \E(\phi(v)) = 0 \quad \mbox{ and } \quad 
\E(\phi(v)\phi(w)) = \la v,w \ra \quad \mbox{ for } \quad v,w \in \cH.\] 
\end{definition}

\begin{remark} \label{rem:rep-mot} 
There are many realizations of Gaussian measure spaces over real 
Hilbert spaces. The one chosen above has the advantage that it 
directly leads to a natural unitary representation of the 
motion group $\Mot(\cH) \cong \cH \rtimes \OO(\cH)$ of $\cH$ 
by 
\[ (\rho(b,g)F)(\alpha) = e^{i \phi(b)(\alpha)} F(g^*\alpha)\] 
for which the map $v \mapsto e^{i\phi(v)} = \widetilde\rho(v,\1)1$ is equivariant 
(cf.\ Remark~\ref{rem:fock-kernel}). Its range is total, and the corresponding 
positive definite function on $\Mot(\cH)$ is given by 
\begin{equation}
  \label{eq:posdef-fun}
 \vphi(b,g) = \la 1, \rho(b,g)1 \ra = \E(e^{i\phi(b)}) = e^{-\|b\|^2/2}.
\end{equation}
\end{remark}

\begin{remark} \label{rem:2.1} Suppose that  $\cH$ is real or complex Hilbert space, $T$ a set,  $\gamma \colon T \to \cH$ a map, and 
$(\phi(v))_{v \in \cH}$ the canonical Gaussian process indexed by $\cH$ 
(Definition~\ref{def:3.5}). Then 
$(\phi(\gamma(t)))_{t \in T}$ is a centered Gaussian process indexed by $T$ with 
covariance kernel 
\[ C(s,t) = \E\big(\overline{\phi(\gamma(s))}\phi(\gamma(t))\big) = \la \gamma(s), \gamma(t)\ra.\] 
In view of Remark~\ref{rem:kerspace}, a kernel on $T$ is the covariance kernel 
of a Gaussian process if and only if it is positive definite. 
If $\gamma(T)$ is total in $\cH$, then the corresponding Gaussian process is 
{\it full} in the sense that, up to sets 
of measure $0$, $\Sigma$ is the smallest $\sigma$-algebra for which 
every $X_t$ is measurable. 
Conversely, for every full and centered Gaussian process $(X_t)_{t \in T}$ with 
covariance kernel $C$  on the probability space 
$(Q,\Sigma,\mu)$, there exists a uniquely determined 
unitary operator \break $U \colon \Gamma(\cH) \to L^2(Q,\Sigma,\mu)$ 
with $U(\phi(\gamma(t))) = X_t$ for $t \in T$ 
(Thm.~1.10 in Ref.~\cite{Hid80}).
\end{remark}

\begin{definition} (Normalization) If $(X_t)_{t \in T}$ is a centered Gaussian process with 
$\E(|X_t|^2) > 0$ for every $t \in  T$ and covariance kernel $C$, then 
$\widetilde X_t := X_t/\sqrt{\E(|X_t|^2)}$ is called the {\it associated normalized 
process}. Its covariance kernel has the form 
\[ \widetilde C(t,s) = \frac{C(t,s)}{\sqrt{C(t,t)C(s,s)}} \quad \mbox{ for } \quad t,s \in T.\]
\end{definition}

\begin{definition} \label{def:b.5}
Let $(X_t)_{t \in T}$ be a centered real-valued Gaussian process 
with covariance kernel $C \colon T \times T\to \R$ 
and $\sigma \colon G \times T \to T, (g,t) \mapsto g.t$ a group action. 

(a) The process $(X_t)_{t \in T}$ is called {\it stationary} 
if 
\[ C(g.t,g.s) = C(t,s) \quad \mbox{ for } \quad g \in G, t,s \in T.\] 
For any realization $\gamma \colon T \to \cH$ with total range as in 
Remark~\ref{rem:2.1}, we thus obtain a uniquely determined orthogonal 
representation $U \colon G \to \OO(\cH)$ satisfying 
$U \gamma(t) = \gamma(g.t)$ for $g \in G, t \in T$. 

(b) The process $(X_t)_{t \in T}$ is said to have {\it stationary increments} 
if the kernel 
\begin{equation}
  \label{eq:d-kern}
D(t,s) := \E\left((X_t - X_s)^2\right) = C(t,t) + C(s,s) - 2 C(t,s), 
\end{equation}
is $G$-invariant. 
For any realization $\gamma \colon T \to \cH$ with total range as in 
Remark~\ref{rem:2.1}, we thus obtain a unique affine action 
$\alpha \colon G\to \Mot(\cH)$ satisfying 
$\alpha_g\gamma(t) = \gamma(g.t)$ for $g \in G, t \in T$ 
\cite{Gu72, HV89}. 
\end{definition}

\begin{remark}
\label{rem:fock-kernel} 
(a) The canonical Gaussian process $(\phi(v))_{v \in \cH}$ over the real Hilbert space 
$\cH$ is stationary for the orthogonal group $\OO(\cH)$ and has stationary increments 
for the motion group $\Mot(\cH)$ because the kernel 
\[ D(v,w) = \E((\phi(v)-\phi(w))^2) = \|v\|^2 + \|w\|^2 - 2 \la v, w\ra 
= \|v - w \|^2 \] 
is invariant under all isometries. 

(b) We also observe that 
\[ K(v,w) 
:= \E(\overline{e^{i\phi(v)}}e^{i\phi(w)}) 
= \E(e^{i\phi(w-v)}) = e^{-\frac{\|v-w\|^2}{2}}.\] 
The functions $(e^{i\phi(v)})_{v \in \cH}$ form a total subset of 
$\Gamma(\cH)$ and the map 
$\eta \colon \cH \to \Gamma(\cH), \eta(v) =  e^{i\phi(v)}$ 
is $\Mot(\cH)$-equivariant with respect to the representation $\rho$.
This reflects the $\Mot(\cH)$-invariance of the kernel $K$. 

(b) The random field $(e^{i\phi(v)})_{v \in \cH}$ arises by normalization of the process given by 
\[ \Gamma(v) := e^{\|v\|^2/2} e^{i\phi(v)} \quad \mbox{ with } \quad 
\E(\overline{\Gamma(v)}\Gamma(w))  = e^{\frac{\|v\|^2}{2} + \frac{\|w\|^2}{2}} e^{-\frac{\|v-w\|^2}{2}}
= e^{\la v, w \ra}.\] 
\end{remark} 

\section{Reflection positive  affine actions} 

Let $(G,S,\tau)$ be a symmetric Lie group 
and $\cE$ be a real Hilbert space, 
endowed with an isometric involution $\theta$. 
We  consider an affine isometric action 
\[ \alpha_g v = U_g v + \beta_g \quad \mbox{ for } \quad g \in G, v \in \cE,\]
where $U \colon G \to \OO(\cE)$ is an orthogonal representation 
and $\beta \colon G \to \cE$ a $1$-cocycle, i.e., 
$\beta_{gh} = \beta_g + U_g \beta_h$ for $g,h \in G.$
We further assume that $\theta \alpha_g \theta = \alpha_{\tau(g)},$ 
which is equivalent to 
\[  \theta U_g \theta = U_{\tau(g)} \quad \mbox{ and } \quad 
\theta \beta_g = \beta_{\tau(g)} \quad \mbox{ for } \quad g \in G.\] 
Then the positive definite kernel 
$C(s,t) := \la \beta_s, \beta_t \ra$ 
and the negative definite function $\psi(g) := \|\beta_g\|^2$ 
on $G$ are related by 
\[ C(s,t) = \frac{1}{2}\big(\psi(s) + \psi(t) - \psi(s^{-1}t)\big), 
\quad 
\psi(s^{-1}t) = C(s,s) + C(t,t) - 2 C(t,s) = \|\beta_s-\beta_t\|^2 \] 
(cf.\ \eqref{eq:d-kern}). 
In particular, the affine action $\alpha$ can be recovered completely from the 
function~$\psi$ (cf.\ Definition~\ref{def:b.5}, \cite{Gu72, HV89}). 
We also note that $\theta\beta_g = \beta_{\tau(g)}$ implies 
that $\psi \circ \tau = \psi$. 
The action of $G$ on $\cE$ preserves the positive definite $\theta$-invariant kernel 
$Q(x,y) := e^{-\|x-y\|^2/2}$ (Remark~\ref{rem:fock-kernel}). 

\begin{definition} (Reflection positive affine actions) 
We now consider the closed subspace $\cE_+$ generated by $(\beta_s)_{s \in S}$ 
and observe that it is invariant under the affine action of $S$ on~$\cE$ 
because $\alpha_s \beta_t = \beta_{st}$ for $s,t \in S$. 
We call the affine action $(\alpha,\cE)$ is {\it reflection positive} if 
$\cE_+$ is $\theta$-positive. 
\end{definition}

The $\theta$-positivity of $\cE_+$ is equivalent to the 
positive definiteness of the kernel 
\[ C^\tau(s,t) =  C(s,\tau(t)) =  C(\tau(s),t)
= \frac{1}{2}\big(\psi(s) + \psi(t) - \psi(s^\sharp t)\big) \]
on~$S$. This in turn is equivalent to the negative definiteness of 
the function $\psi\res_S$ (Definition~\ref{def:8.1.1}). 
Then $e^{-\psi(s)/2}$ is positive definite on $S$ and the Gelfand--Naimark--Segal 
construction leads to the corresponding contraction representation 
(Lem.~1.4, Prop.~1.11(ii) in Ref.~\cite{NO14}). 
This leads us to the following concept: 

\begin{definition} We call a continuous function $\psi \colon G \to \R$ {\it reflection negative} 
(with respect to $(G,S,\tau)$) if 
$\psi$ is a negative definite function on $G$ 
and $\psi\res_S$ is a negative definite function on the involutive semigroup 
$(S,\sharp)$ (Definition~\ref{def:8.1.1}). 
\end{definition}

\begin{remark} According to Schoenberg's Theorem for kernels (Thm.~3.2.2 in 
Ref.~\cite{BCR84}), 
a function $\psi \colon G \to \C$ is 
reflection negative if and only if, for every $\lambda > 0$, the function 
$e^{-\lambda \psi}$ is reflection positive\cite{NO14}. 
\end{remark}

{}From now on we focus on some specific aspects of the 
special case $(G,S,\tau)= (\R,\R_+,-\id)$. 
Recall that a smooth function $\psi \colon (0,\infty) \to \R$ is called a {\it Bernstein function} 
if $\psi \geq 0$ and ${(-1)^{k-1} \psi^{(k)} \geq 0}$ for every $k \in \N$. 
Combining Cor.~3.3 in Ref.~\cite{NO14} with Bernstein's Theorem (Thm.~3.2 in 
Ref.~\cite{SSV10}) we 
obtain: 

\begin{theorem} \label{thm:lk-bernstein}
A symmetric continuous function $\psi \colon \R \to \R$ is reflection negative if and only 
if $\psi\res_{(0,\infty)}$ is a Bernstein function. In particular, this is equivalent to 
the existence of $a,b \geq 0$ and a positive measure $\mu$ on $(0,\infty)$ 
with $\int_0^\infty (1 \wedge \lambda)\, d\mu(\lambda) < \infty$ 
such that 
\[ \psi(t) = a + b|t| + \int_0^\infty (1 - e^{-\lambda |t|})\, d\mu(\lambda) \] 
(L\'evy--Khintchine representation). Here $a,b$ and $\mu$ are uniquely determined by~$\psi$.
\end{theorem}

\begin{example} \label{ex:bernstein} 
(Cor.~3.2.10 in Ref.~\cite{BCR84}) For $\alpha \geq 0$, the function 
$\psi(t) := |t|^\alpha$ is reflection negative on $\R$ if and only if 
$0 \leq \alpha \leq 1$. For $0 < \alpha < 1$, this follows from 
the integral representation 
$t^\alpha = \frac{\alpha}{\Gamma(1-\alpha)} \int_0^\infty (1- e^{-\lambda t}) 
\lambda^{-1-\alpha}\, d\lambda$ for $t > 0$. 
\end{example}

\section{Relations to stochastic processes} 

In this last section, we discuss the abstract concepts from above in the 
context of natural Gaussian processes on the real line, such as Brownian motion 
and its relatives.

\begin{example} {\it Two sided Brownian motion} is a real-valued centered Gaussian process 
$(B_t)_{t \in \R}$ with covariance kernel 
\[ C(s,t) = \E(B_s B_t) = \frac{1}{2}(|s| + |t| - |s-t|) \quad \mbox{ for } \quad 
s,t \in \R\] 
and 
\[ D(s,t) = \E((B_s-B_t)^2) = C(s,s) + C(t,t) - 2 C(s,t) = |s-t|.\] 
In particular, it has stationary increments. A natural realization in 
$\Gamma(L^2(\R))$ is given by 
$B_t = \phi(b_t)$ for $b_t := \sgn(t) \chi_{[t \wedge 0, t \vee 0]}.$
Then the corresponding affine isometric action of the additive group 
$\R$ on $L^2(\R)$ is given by 
\[ \alpha_t f = S_t f + b_t, \quad \mbox{ where } \quad (S_t f)(x) = f(x-t).\] 
This action is reflection positive because the function 
$\psi(t) := \|b_t\|^2 = |t|$ is reflection negative for $(\R,\R_+,-\id_\R)$.
Since $C(s,-t) = 0$ for $t,s \geq 0,$ 
we get $\widehat\cE=\{0\}$ if $\cE = L^2(\R)$ and 
$\cE_+ = L^2(\R_+)$. 
 
 More generally,  for a reflection positive affine action 
$(\alpha,\cE)$ of $\R$ one can show that $\widehat\cE$ is trivial if and only if 
$\la\beta_s, \beta_t \ra = 0$ for $ts < 0$ (see Ref.~\cite{JNO16}). This property characterizes Brownian motion (up to positive multiples) 
among  Gaussian processes with stationary increments. 
\end{example}

\begin{example} For $0 < H < 1$, 
fractional Brownian motion is a centered Gaussian process $(X_t)_{t \in \R}$ 
with stationary increments, covariance kernel 
\[ C(t,s) = \frac{1}{2}(|t|^{2H} + |s|^{2H} - |t-s|^{2H}), \quad 
D(t,s) = C(t,t) + C(s,s) - 2 C(t,s) = |t-s|^{2H}.\]  
Therefore it corresponds to the negative definite function 
$\psi(t) := |t|^{2H}$, which is reflection negative if and only if 
$0 \leq H \leq \frac{1}{2}$ (Example~\ref{ex:bernstein}). 
Note that $H= 1/2$ corresponds to Brownian motion. 
\end{example}
 
 \begin{example} One sided Brownian motion $(B_t)_{t > 0}$ has the covariance 
kernel $C(s,t) = s \wedge t$ and the corresponding normalized process on 
$(0,\infty)$ has the kernel 
$\widetilde C(s,t) = \frac{s \wedge t}{\sqrt{st}} = \sqrt{\frac{s \wedge t}{s \vee t}}.$ 
This kernel is invariant under the dilation group $\R^\times_+$, so that 
the process 
$(\widetilde B_t)_{t > 0}$ is stationary with respect to  dilations. 

Accordingly, the natural realization by $B_t = \phi(b_t)\in 
\Gamma(L^2(\R_+))$ with $b_t = \chi_{[0,t]}$ and its normalization 
$\widetilde b_t = t^{-1/2} b_t$ are compatible with the unitary  
representation of the dilation group on $L^2(\R_+)$ 
by $(\tau_t f)(x) = e^{t/2} f(e^{t}x)$ in the sense that 
$\widetilde b_{e^{-t}} = \tau_t \chi_{[0,1]} = \tau_t \widetilde b_1$. 

The invariance of the kernel $\widetilde C$ under the involution 
$\tau(t) = t^{-1}$ of $\R^\times$ implies on $L^2(\R_+)$ 
the existence of a unique unitary involution $\theta$ 
with $\theta(\widetilde b_t) = \widetilde b_{1/t}$. 
On $(0,1)$, the reflected kernel 
$\widetilde C(s,t^{-1}) 
=\sqrt{\frac{s \wedge t^{-1}}{s \vee t^{-1}}}
=\sqrt{\frac{s}{t^{-1}}} = \sqrt{st}$ 
is positive definite, so that we obtain 
with $\cE = L^2(\R_+)$ and $\cE_+ = L^2([0,1])$ 
a reflection positive Hilbert space on which the dilation 
representation of $\R$ is reflection positive. 
The corresponding reflection positive function on $\R$ is 
\[ \vphi(t) 
= \la \chi_{[0,1]}, \tau_t \chi_{[0,1]} \ra
= \la \widetilde b_1, \widetilde b_{e^{-t}} \ra = \widetilde C(1,e^{-t}) 
= e^{-|t|/2}\]
and $\widehat\cE$ is one dimensional.
\end{example}

\section*{Acknowledgments} 

The research of P.~Jorgensen was partially supported by the
Binational Science Foundation Grant number 2010117. 
K.~H.~Neeb was supported by DFG-grant NE 413/7-2, Schwerpunktprogramm 
``Darstellungstheorie'' and 
Gestur \'Olafsson by NSF grant  DMS-1101337.


\begin{thebibliography}{10}

\bibitem{GJ81} J. Glimm and A. Jaffe, {\it Quantum Physics--A Functional Integral 
Point of View} (Springer-Verlag, New York, 1981)

\bibitem{Kl77} A. Klein,  Gaussian OS-positive processes,
{\it Z. Wahrscheinlichkeitstheorie und Verw. Gebiete} {\bf 40:2}, 115 (1977)

\bibitem{KL82} A. Klein  and L. Landau, From the Euclidean group to the
Poincar\'e group via Osterwalder-Schrader positivity, {\it Comm. Math. Phys.} 
{\bf 87}, 469 (1982/83)

\bibitem{JO00} P. E. T. Jorgensen, and G. \'Olafsson, 
 Unitary representations and Osterwalder-Schrader
duality, in {\it The Mathematical Legacy of Harish--Chandra,} 
R. S. Doran and V. S. Varadarajan, eds., Proc. Symp. in Pure Math. {\bf 68} 
(Amer. Math. Soc., 2000) 

\bibitem{JR07} A. Jaffe and G. Ritter,  Quantum field theory 
on curved backgrounds. I. The euclidean functional integral,
{\it Comm. Math. Phys.} {\bf 270}, 545 (2007).

\bibitem{OS73} K. Osterwalder and R. Schrader, Axioms for Euclidean Green's functions.~1,
{\it Comm. Math. Phys.} {\bf 31}, 83 (1973)

\bibitem{NO14} K.-H. Neeb and  G. \'Olafsson,  Reflection 
positivity and conformal symmetry, {\it J. Funct. Anal.} {\bf 266}, 2174 (2014)

\bibitem{MNO14} S. Merigon, K.-H. Neeb and G. \'Olafsson, 
Integrability of unitary representations on reproducing kernel spaces, 
{\it Representation Theory} {\bf 19}, 24 (2015)

\bibitem{LM75} M. L{\"u}scher and G. Mack, Global conformal invariance in quantum field theory, {\it Comm. Math. Phys.} \textbf{41}, 203 (1975)

\bibitem{JNO16} P. E. T. Jorgensen, K.-H. Neeb, and G. \'Olafsson, 
Reflection negative kernels and relations to stochastic processes, 
in preparation

\bibitem{BCR84} C. Berg, J. P. R. Christensen, and P. Ressel, {\it Harmonic Analysis on Semigroups} (Graduate Texts in Math., Springer-Verlag, Berlin, Heidelberg, 1984)

\bibitem{Hid80} T. Hida, {\it Brownian Motion,}
Applications of Mathematics {\bf 11} (Springer, 1980)

\bibitem{JNO15} P. E. T. Jorgensen, K.-H. Neeb, and G. \'Olafsson, 
Reflection positive stochastic processes indexed by Lie groups, 
arXiv:math-ph:1510.07445

\bibitem{Si74} B. Simon, {\it The $P(\Phi)_2$ Euclidean (Quantum) Field Theory}, 
(Princeton Univ. Press, 1974) 

\bibitem{Gu72} A. Guichardet, {\it Symmetric Hilbert spaces and related topics. 
Infinitely divisible positive definite functions. 
Continuous products and tensor products. Gaussian and Poissonian stochastic processes},
Lecture Notes in Mathematics {\bf 261} (Springer-Verlag, Berlin-New York, 1972)

\bibitem{HV89} P.~de la Harpe and  A. Valette, 
{\it La propri\'et\'e $(T)$ de Kazhdan pour les groupes localement 
compacts} Ast\'erisque  {\bf 175} (Soc. Math. de France, 1989)

\bibitem{SSV10} R. Schilling, R. Song and Z. Vondracek, {\it Bernstein Functions,}
Studies in Math. (de Gruyter, 2010) 

\end{thebibliography}
\end{document}